\documentclass[a4paper]{revtex4}
\usepackage{graphicx}
\graphicspath{{figures/}}
\usepackage{fancyhdr}
\usepackage{amsmath}
\usepackage{color}

\begin{document}

\title{hhh Coupling in SUSY models after LHC run I}

%
\author{Lei Wu}
\affiliation{ARC Centre of Excellence for Particle Physics at the Terascale, School of Physics, The University of Sydney, NSW 2006, Australia}
\author{Jin Min Yang}
\affiliation{State Key Laboratory of Theoretical Physics, Institute of Theoretical Physics, Academia Sinica, Beijing 100190, China}
\author{Chien-Peng Yuan}
\affiliation{Department of Physics and Astronomy, Michigan State University, East Lansing, Michigan 48824, USA}
\author{Mengchao Zhang}
\affiliation{State Key Laboratory of Theoretical Physics, Institute of Theoretical Physics, Academia Sinica, Beijing 100190, China}

\begin{abstract}
We examine the Higgs triple coupling in MSSM and NMSSM under current constraints, which include the LHC measurements, Higgs data, B physics, electroweak precision observables, relic density and so on.
The ratio $\lambda^{\rm MSSM}_{hhh}/\lambda^{\rm SM}_{hhh}$ is above 0.97, due to the highly constrained parameter space.
While the ratio $\lambda^{\rm NMSSM}_{hhh}/\lambda^{\rm SM}_{hhh}$ can reach 0.1 under current constraints.
The precise measurement in future collider will give a tighter constraint to the Higgs triple coupling in MSSM and NMSSM.
\end{abstract}

\maketitle

\thispagestyle{fancy}


\section{Introduction}
The Standard Model(SM) spectrum have been completed by the discovery of 125GeV Higgs\cite{atlas,cms}. But in order to describe the Higgs potential and pin down the SM, we still need to know the Higgs self-couplings.
In New Physics Models which contain more than one Higgs, the Higgs self-coupling will change a lot.
And such an different self-coupling is very important in many theory and phenomenology problem, such as baryon-geneses and vacuum stability.
So it's important to calculate how large this change can be under current constraints.

An interesting feature in the MSSM Higgs sector have been shown in \cite{feyn-mssm1}. It says that even the stop loop correction to the Higgs self coupling is significant, but such a large New Physics contribution can always be absorbed by the redefinition of the Higgs mass. So, after a detailed matching of the Higgs mass and Higgs self coupling, the MSSM will show decoupling property, which means all the things you can see in low energy scale is just the SM when all of the non-SM particles are very heavy. Such a matching between Higgs mass and its coupling can be automatically obtained in the effective potential method \cite{eff1}\cite{eff2}\cite{eff3}\cite{eff4}. So as suggested by \cite{james}, we will use the effective potential method to perform our calculation.
\section{Higgs sector in MSSM and NMSSM}
The MSSM Higgs sector have two doublets:
\begin{eqnarray}
H_u = \left(\begin{array}{c} H_u^+ \\ H_u^0 \end{array} \right), \qquad
H_d = \left(\begin{array}{c} H_d^0 \\ H_d^- \end{array} \right) \, .
\end{eqnarray}
The tree-level Higgs potential can be expressed as:
\begin{eqnarray}
V^{(0,MSSM)} & = & m_1^2 |H_u|^2 + m_2^2 |H_d|^2 - B_\mu \epsilon_{\alpha\beta} (H_u^\alpha H_d^\beta
+ h.c.) \nonumber \\
& & +\frac{g^2+g'^2}{8} (|H_u|^2-|H_d|^2)^2 + \frac{g^2}{2}
|H_u^\dagger H_d|^2 \; ,
\end{eqnarray}
Using the minimum condition, the soft mass $m_1^2$, $m_2^2$, and $ B_\mu$ can be expressed by $M_A$, $\tan\beta$ and other SM parameters. So the tree level MSSM Higgs sector can be decided by $M_A$ and $\tan\beta$. All the couplings in the tree-level MSSM Higgs sector are gauge couplings, and the stop/top sector will give a large quantum correction to tree-level potential.

Beside the two Higgs doublets, the NMSSM Higgs sector is extended by a singlet $\langle S$, and a effective $\mu$ term will be introduced by the spontaneous symmetry breaking of this singlet field.
The tree-level Higgs potential in NMSSM is:
\begin{eqnarray}
V^{(0,NMSSM)} & = & (|\lambda S|^2 + m_{H_u}^2)H_{u}^\dagger H_{u} + (|\lambda S|^2 + m_{H_d}^2)H_{d}^\dagger H_{d}
 +m_S^2 |S|^2 \nonumber \\
& &  + \frac{1}{8} (g_2^2+g_1^{2})( H_{u}^\dagger H_{u}-H_{d}^\dagger H_{d} )^2+\frac{1}{2}g_2^2|H_{u}^\dagger H_{d}|^2
 + | \lambda  H_{u}  H_{d} + \kappa S^2 |^2+ \nonumber \\
& & \big[\lambda A_\lambda H_{u}  H_{d} S  +\frac{1}{3} \kappa
A_{\kappa} S^3+h.c. \big] \,,
\end{eqnarray}
with $\kappa$ and $\lambda$ are dimensionless parameters, and $A_{\lambda}$ and $A_{\kappa}$ are dimensional trilinear terms. The largest different between MSSM and NMSSM Higgs sector is the Higgs coupling(tree-level).
As we mentioned before, the Higgs coupling in MSSM are all gauge couplings, while the Higgs coupling is NMSSM is very different, they only obey some theoretical constraints like perturbativity in high energy scale, vaccum stability, or the phenomenology constraints from Higgs data. So the Higgs triple coupling in NMSSM can be very different with SM and MSSM. Next we will show the exact number of Higgs triple coupling in MSSM and NMSSM under current constraints.

\section{Numerical calculations and results}
The input SM parameters are \cite{pdg}:
\begin{eqnarray}
&&m_t=173.5{\rm ~GeV}, \quad m_{W}=80.385{\rm ~GeV}, \quad m_{Z}=91.1876 {\rm ~GeV},\nonumber\\
&&m^{\overline{MS}}_{b}(m^{\overline{MS}}_{b})=4.18{\rm ~GeV}, \quad \alpha_s(M_Z)=0.1184, \quad \alpha(m_Z)^{-1}=128.962
\end{eqnarray}
The mass of sleptons, first and second generation squarks, gaugino are all set to 2TeV. 
We further require $M_{Q3}=M_{U3}=M_{D3}$ and $A_t=A_b$.
Then we use NMSSMTools-4.4.1\cite{NMSSMTOOLS} to perform a random scan. For MSSM(MSSM can be treated as a limiting scenario of NMSSM, so NMSSMTools can be used for the MSSM scan), the scan range is:
\begin{eqnarray}
&&1 \le \tan\beta \le 60, ~ 0.2 {~\rm TeV} \le \ M_A \le 1 {~\rm TeV},
~|\mu| \le 1 {~\rm TeV}, \nonumber \\
&&
0.1 {~\rm TeV} \le M_{Q3} \le 2.5 {~\rm TeV}, ~ |A_t| \le 3M_{Q3}.
\end{eqnarray}
And the scan range of NMSSM is:
\begin{eqnarray}
&&0 \le \lambda \le 0.7, ~ |\kappa| \le 0.7,
 ~0.2 {~\rm TeV} \le A_\lambda \le 1 {~\rm TeV},  \nonumber\\
&&
|A_\kappa| \le 1 {~\rm TeV}, ~1 \le \tan\beta \le 20,  ~~0.1 {~\rm TeV} \le \mu \le 1 {~\rm TeV}, \nonumber\\
&& 0.1 {~\rm TeV} \le M_{Q3} \le 1 {~\rm TeV}, ~~|A_t| \le 3M_{Q3}.
\end{eqnarray}

The included constraints are:
\begin{itemize}
  \item Higgs data. HiggsBounds-4.2.\cite{higgsbounds} and HiggsSignals-1.3.0\cite{higgssignals} have been used to exclude the points that out of the 2$\sigma$ region.
  \item B physics observables $B\to X_s\gamma$, $B_s\to \mu^+\mu^-$, $B_d\to X_s\mu^+\mu^-$ and $ B^+\to \tau^+\nu$ are required to be in 2$\sigma$ region.
  \item Precision electroweak observables are required to be in 2$\sigma$ region.
  \item Relic density required to be below the 2$\sigma$ upper bound of the Planck value\cite{planck}. The neutralino-proton scattering cross section are required to satisfy the direct detection bound from LUX\cite{lux}.
  \item The physical vacuum should be stable.
\end{itemize}

In order to see the new physics effect, it's will be convenient show the ratio $\lambda^{\rm (N)MSSM}_{hhh}/\lambda^{\rm SM}_{hhh}$.
If this ratio is unequaled to one, then we will see a SUSY effect. And the farther this ratio deviate one, the larger SUSY effect we can see.

\begin{figure}[ht]
\centering
\includegraphics[width=5in]{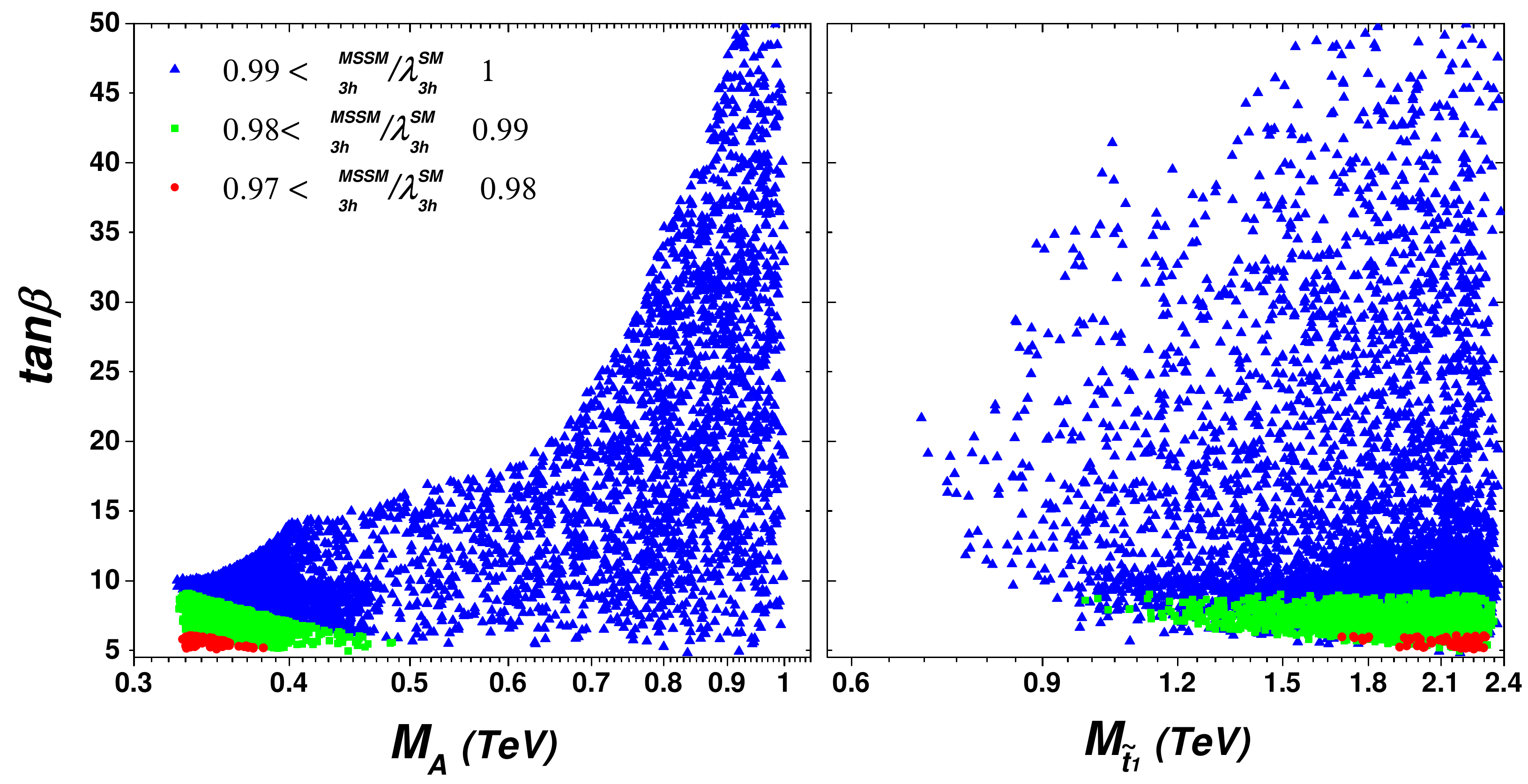}\vspace{-0.2cm}
\caption{Scatter plots of the MSSM samples surviving all the experimental constraints,
projected on the planes of $\tan\beta$ versus $m_{A}$ and $m_{\tilde{t}_{1}}$.}
\label{mssm-3h}
\end{figure}

\begin{figure}[ht]
\centering
\includegraphics[width=5in]{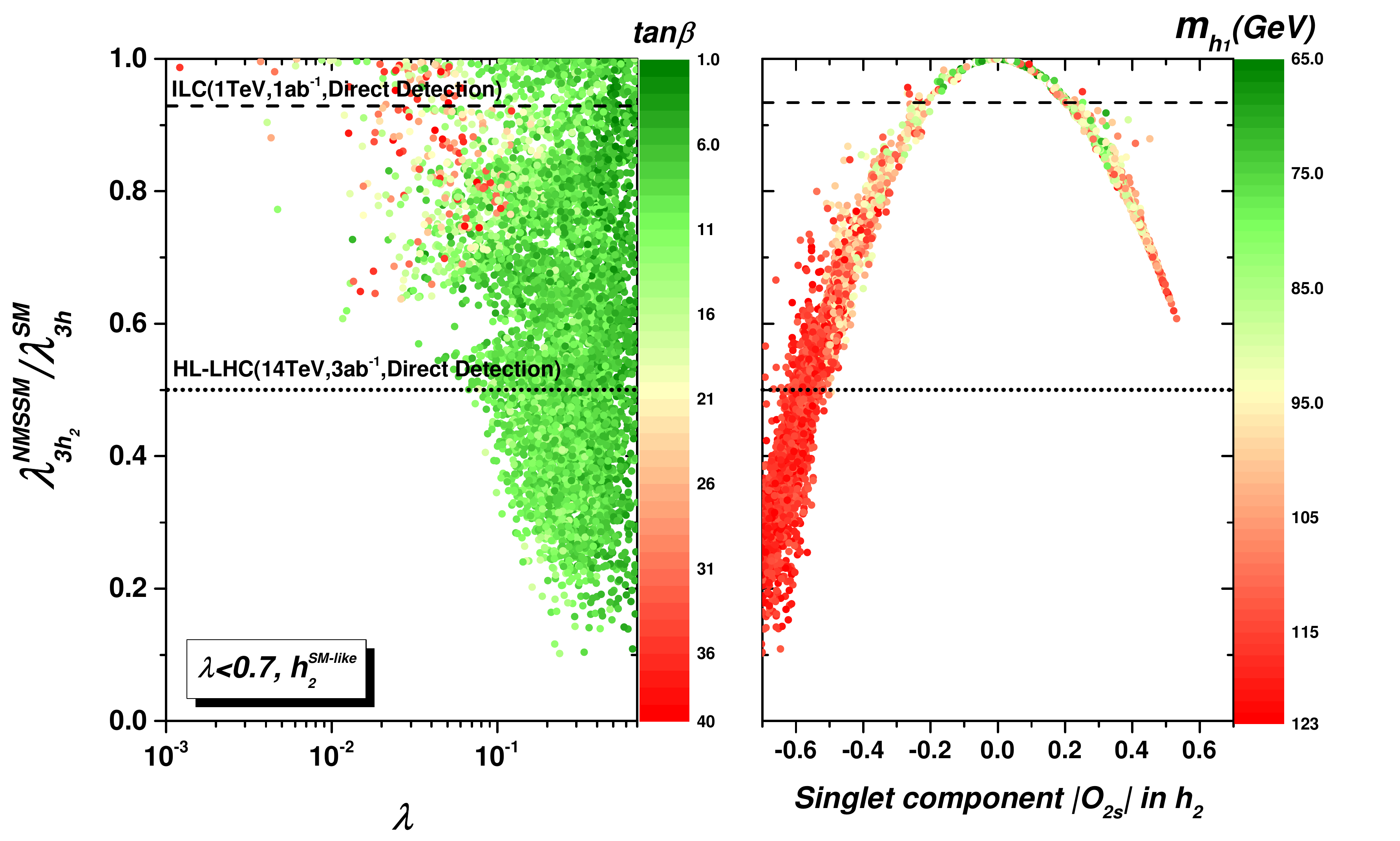}\vspace{-0.2cm}
\caption{The dependence of $\lambda^{NMSSM}_{3h_2}/\lambda^{SM}_{3h}$ versus $\lambda$ and the singlet component $|{\cal O}_{2s}|$ in $h_2$, where $m_{h_2} < 2 m_{h_1}$. The ILC(1 TeV, 1 ab$^{-1}$) and HL-LHC(14 TeV, 3 ab$^{-1}$) sensitivities are also plotted (the region below each horizontal line is detectable).}
\label{nmssm-3h}
\end{figure}

Fig.\ref{mssm-3h} and Fig.\ref{nmssm-3h} are our main scan results. The Higgs triple coupling deviation in MSSM is to small to be seen in future colliders, while this triple coupling deviation in NMSSM is larger enough to be detected by future colliders. And the large triple coupling deviation in NMSSM always correspond to a scalar in the neighbour of the 125GeV Higgs. Detailed discussion can be found in our longer paper \cite{Wu:2015nba}.

\section{the Constraint to hhh Coupling in Future Collider}
\begin{figure}[ht]
\centering
\includegraphics[width=5in]{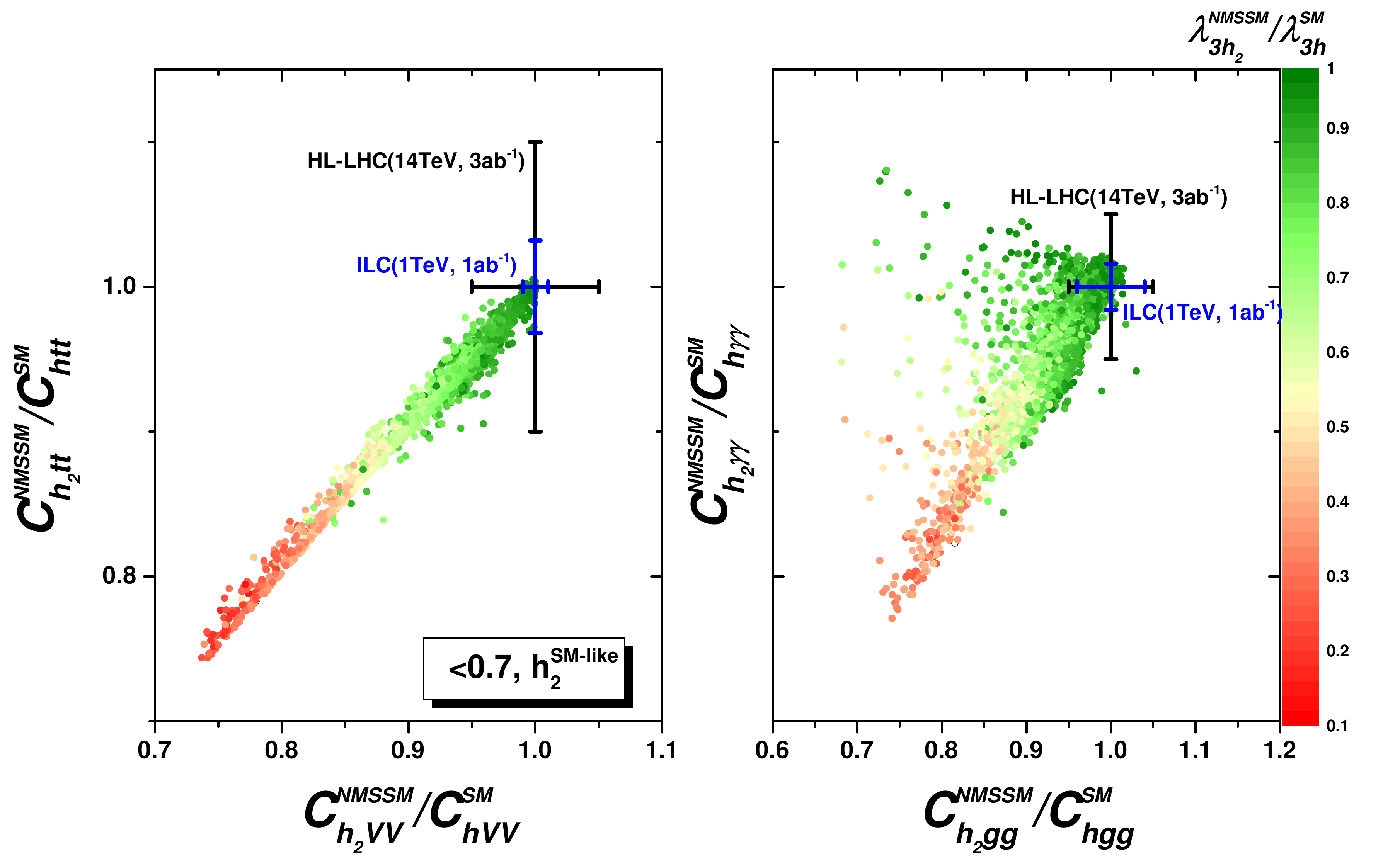}\vspace{-0.2cm}
\caption{The relation of $hVV$, $hf\bar{f}$ and Higgs triple coupling in NMSSM. Crosses are the Higgs coupling measurement error bar in future colliders.}
\label{nmssm-couplings}
\end{figure}

The large deviation of Higgs triple coupling will happen when the SM-like Higgs mix with other non-SM scalar adequately. 
But such a adequate mixing will also cause a deviation in $hVV$, $hf\bar{f}$. 
So if we could measure the $hVV$, $hf\bar{f}$ coupling much precisely, the constraint to Higgs triple coupling will be more stringent. Here we just show the NMSSM result in Fig.\ref{nmssm-couplings}.

\section{Conclusion}
In this talk we show the Higgs triple coupling deviation is MSSM and NMSSM under current constraints. 
The ratio $\lambda^{\rm MSSM}_{hhh}/\lambda^{\rm SM}_{hhh}$ is above 0.97 and the ratio $\lambda^{\rm NMSSM}_{hhh}/\lambda^{\rm SM}_{hhh}$ can reach 0.1 under current constraints. The precise Higgs coupling measurement in future collider can constraint the Higgs triple coupling deviation tighter.

\section*{Acknowledgement}
Thanks Roman Nevzorov, Ulrich Ellwanger, Archil Kobakhidze and Michael Schmidt for very helpful discussions. This work was partly supported by the Australian Research Council, by the CAS Center for Excellence in Particle Physics (CCEPP), the National Natural Science Foundation of China (NNSFC) under grants Nos. 11305049, 11405047, 11275245, 10821504 and 11135003, by Specialized Research Fund for the Doctoral Program of Higher Education under Grant No.20134104120002, and by the U.S. National Science Foundation under Grant No. PHY-1417326.

\bigskip 

\end{document}